\documentclass[12pt]{iopart}
\usepackage[colorlinks=true,citecolor=blue,linkcolor=blue,urlcolor=blue]{hyperref}
\usepackage{amssymb}
\usepackage{graphicx}
\usepackage[compress]{cite}
\usepackage[usenames]{color}
\usepackage{subfigure}
\usepackage{epsfig}

\newcommand{\ket}[1]{\vert #1 \rangle}
\newcommand{\bra}[1]{\langle #1 \vert}

\newcommand{\abs}[1]{| #1 |}

\begin{document}

\title{Analog Quantum Approximate Optimization Algorithm} 

\author{Nancy Barraza$^{1}$, Gabriel Alvarado Barrios$^{1}$, Jie Peng$^2$, Lucas Lamata$^{3,4}$, Enrique Solano$^{1,5,6}$, Francisco Albarr\'an-Arriagada$^{7,8}$}

\address{$^{1}$ International Center of Quantum Artificial Intelligence for Science and Technology (QuArtist) and Department of Physics, Shanghai University, 200444 Shanghai, China}
\address{$^{2}$ Hunan Key Laboratory for Micro-Nano Energy Materials and Devices and School of Physics and Optoelectronics, Xiangtan University, Hunan 411105, China}
\address{$^3$Departamento de F\'isica At\'omica, Molecular y Nuclear, Universidad de Sevilla, 41080 Sevilla, Spain}
\address{$^4$ Instituto Carlos I de F\'isica Te\'orica y Computacional, 18071 Granada, Spain}
\address{$^5$ IKERBASQUE, Basque Foundation for Science, Plaza Euskadi 5, 48009 Bilbao, Spain}
\address{$^6$ Kipu Quantum, Kurwenalstrasse 1, 80804 Munich, Germany}
\address{$^7$ Departamento de F\'isica, Universidad de Santiago de Chile (USACH), Avenida Victor Jara 3493, 9170124, Santiago, Chile}
\address{$^8$ Center for the Development of Nanoscience and Nanotechnology, 9170124, Estaci\'on Central, Chile}

\ead{\mailto{enr.solano@gmail.com}, \mailto{francisco.albarran@usach.cl}}

\begin{abstract}
We present an analog version of the quantum approximate optimization algorithm suitable for current quantum annealers. The central idea of this algorithm is to optimize the schedule function, which defines the adiabatic evolution. It is achieved by choosing a suitable parametrization of the schedule function based on interpolation methods for a fixed time, with the potential to generate any function. This algorithm provides an approximate result of optimization problems that may be developed during the coherence time of current quantum annealers on their way towards quantum advantage.
\end{abstract}

\maketitle

\section{Introduction}
In nature, the dynamics of several relevant systems can be derived from the solution of an optimization problem. Consequently, the development of efficient optimization algorithms has been a central field for computer science, and naturally, this interest in optimization algorithms also arises in quantum computing. One of the most important approaches for solving optimization problems is employing quantum annealers, with remarkable advances by D-Wave company~\cite{Johnson2010SST, Johnson2011Nature}. In this paradigm, a quantum system is adiabatically driven from an initial Hamiltonian (at time $t=0$), with a ground state that is easy to prepare, to a final Hamiltonian (at time $t=T$), whose ground state codifies the solution of the optimization problem. Suppose the evolution time ($T$) is sufficiently large. In that case, the adiabatic theorem ensures that during the evolution, the system will be in the ground state of the instantaneous Hamiltonian and then in the ground state of the final Hamiltonian after the evolution~\cite{Albash2018RMP, Hauke2020RPP}. This approach has been explored for a large variety of problems, from quantum chemistry~\cite{Streif2019QTOP,Genin2019arXiv,Teplukhin2020SciRep} to quantum finance~\cite{Ding2019arXiv} and machine learning~\cite{Willsch2020CPC,Dixit2021FiP}. Nevertheless, the adiabatic evolution demands a large execution time. This time is beyond the coherence time for current quantum annealers, which turns the process incoherent; thus, the possibility of reaching quantum advantage is unclear.

On the other hand, hybrid quantum-classical algorithms have received significant attention during the past few years due to the possibility of being implemented in current noisy intermediate-scale quantum (NISQ) devices. These algorithms are focused on the minimization of a cost function. The cost function is codified in the expectation value of quantum observables, including the Hamiltonian, computed by a quantum processor using parametrized quantum states prepared via parametrized quantum gates. Finally, the minimization is obtained using a classical optimization algorithm over the parameters of the quantum gates. Some examples of these classes of algorithms are the variational quantum algorithms~\cite{Peruzzo2014NatCommun, McClean2016NJP, Cerezo2020arXiv}, the digital quantum approximate optimization algorithm (QAOA)~\cite{Farhi2014arXiv, Hadfield2019Algorithm}, the adaptive random quantum eigensolver \cite{Barraza2021arXiv}, and the digitized counterdiabatic QAOA~\cite{Chandarana2021arXiv} among others. This hybrid approach has been fruitful in different areas, from machine learning to quantum chemistry problems~\cite{Romero2021AQT, Skolik2021arXiv, Chen2020IEEE, Li2019ATS, Kandala2017Nature, Higgott2019Quantum}. Recently, several works have developed faster versions of the hybrid algorithms or their implementation in another paradigm beyond gate-based computing, such as variational algorithms in measured-based quantum computing and some fully-quantum algorithms for optimization problems~\cite{Ferguson2021PRL, Wei2020Research, Wang2019PRL}.
 
 In this work, we propose an analog version of QAOA by the suitable parametrization of a stepwise schedule function followed by a classical optimization. This algorithm can be implementable in current quantum annealers, finding optimal protocols for coherent evolution of the annealer quantum processor and exploiting all the potential of such devices.
 
\section{Analog QAOA}
First of all, an adiabatic algorithm refers to the time evolution given by the following time-dependent Hamiltonian
\begin{equation}
H(t)=[1-\lambda(t/T)]H_{i}+\lambda(t/T)H_{f} \ ,
\label{Eq01}
\end{equation}
where $H_i$ is the initial Hamiltonian with a ground state that is easy to prepare, $H_f$ is the problem Hamiltonian which ground state contains the solution of the optimization problem, $T$  is the total evolution time, and $\lambda(x)$ is the schedule function with $\lambda(0)=0$ and $\lambda(1)=1$. The time evolution is given by ($\hbar=1$ for the rest of the article)
\begin{equation}
U(t)=\mathcal{T}e^{-i\int_0^t\{[1-\lambda(\tau/T)]H_{i}+\lambda(\tau/T)H_{f}\}d\tau} \, ,
\label{Eq02}
\end{equation}
with $\mathcal{T}$ is the time ordering operator. We can digitalize Eq. (\ref{Eq02}) obtaining
\begin{equation}
U(t)\approx e^{-i\sum_{k=0}^N\{[1-\lambda(\delta t_k/2)]H_{i}+\lambda(\delta t_k )H_{f}\}\Delta t_k}\, ,
\end{equation}
where $\delta t_k=(t_{k+1}+t_k)/T$, $\Delta t_k=t_{k+1}-t_k$ and $t_k<t_{k+1}<t_N=T$. Here we have used the discrete-time approximation, where we approximate the continuous time evolution by constant time steps for the schedule function. We can see that $\lambda(\delta t_k/2)$ is the schedule function evaluated in the middle time between $t_k/T$ and $t_{k+1}/T$ which remains constant during the elapsed time $\Delta t_k$. Using the first order Trotter expansion, we have
\begin{eqnarray}
U(t)&&\approx \prod_{k=1}^Ne^{-i\{[1-\lambda(\delta t_k/2)]\Delta t_k H_{i}+\lambda(\delta t_k/2)H_{f}\}\Delta t_k}\nonumber\\
&&\approx\prod_{k=1}^N\left(e^{-i[1-\lambda(\delta t_k/2)]\Delta t_kH_{i}}e^{-i\lambda(\delta t_k/2)\Delta t_kH_{f}}\right).
\label{Eq04}
\end{eqnarray}
This approach called digitized adiabatic quantum computing is suitable to be implemented in gate-based quantum computers \cite{Barends2016Nature}. Notice that each digital step depends on the value of the schedule function $\lambda(\delta t_k/2)$ at that time and the value of the time step $\Delta t_k$. In quantum annealers, $H_i$ is called the mixer Hamiltonian and has the form $H_i=\sum_j\omega_j\sigma_j^x$, while $H_f$ is diagonal in the computational basis in general. 
 
On the other hand, the digital QAOA algorithm is given by a unitary evolution parametrized by $2N$ real numbers
\begin{equation}
\mathcal{U}(\vec{\alpha},\vec{\beta})=\prod_{k=1}^N\left(e^{-i\alpha_kH_{i}}e^{-i\beta_kH_{f}}\right) \, .
\label{Eq05}
\end{equation}
Here, $\vec{\alpha}=\{\alpha_1,\dots,\alpha_N\}$, and $\vec{\beta}=\{\beta_1,\dots,\beta_N\}$ are the parameters to optimize with the cost function
\begin{equation}
\bra{\phi}\mathcal{U}(\vec{\alpha},\vec{\beta})^{\dagger}H_f\mathcal{U}(\vec{\alpha},\vec{\beta})\ket{\phi},
\end{equation}
where $\ket{\phi}$ is the ground state of $H_i$. From Eqs. (\ref{Eq04}) and (\ref{Eq05}), we can see that the digital version of QAOA is basically an optimization of the schedule function $\lambda(x)$ in its digital form, where $\alpha_k=[1-\lambda(\delta t_k/2)]\Delta t_k$ and $\beta_k=\lambda(\delta t_k/2)\Delta t_k$, it means, the optimization of the values $\lambda(\delta t_k/2)$ of the schedule function, as well as the time steps $\Delta t_k$. 

\begin{figure}[t]
\centering
\includegraphics[width=0.8\linewidth]{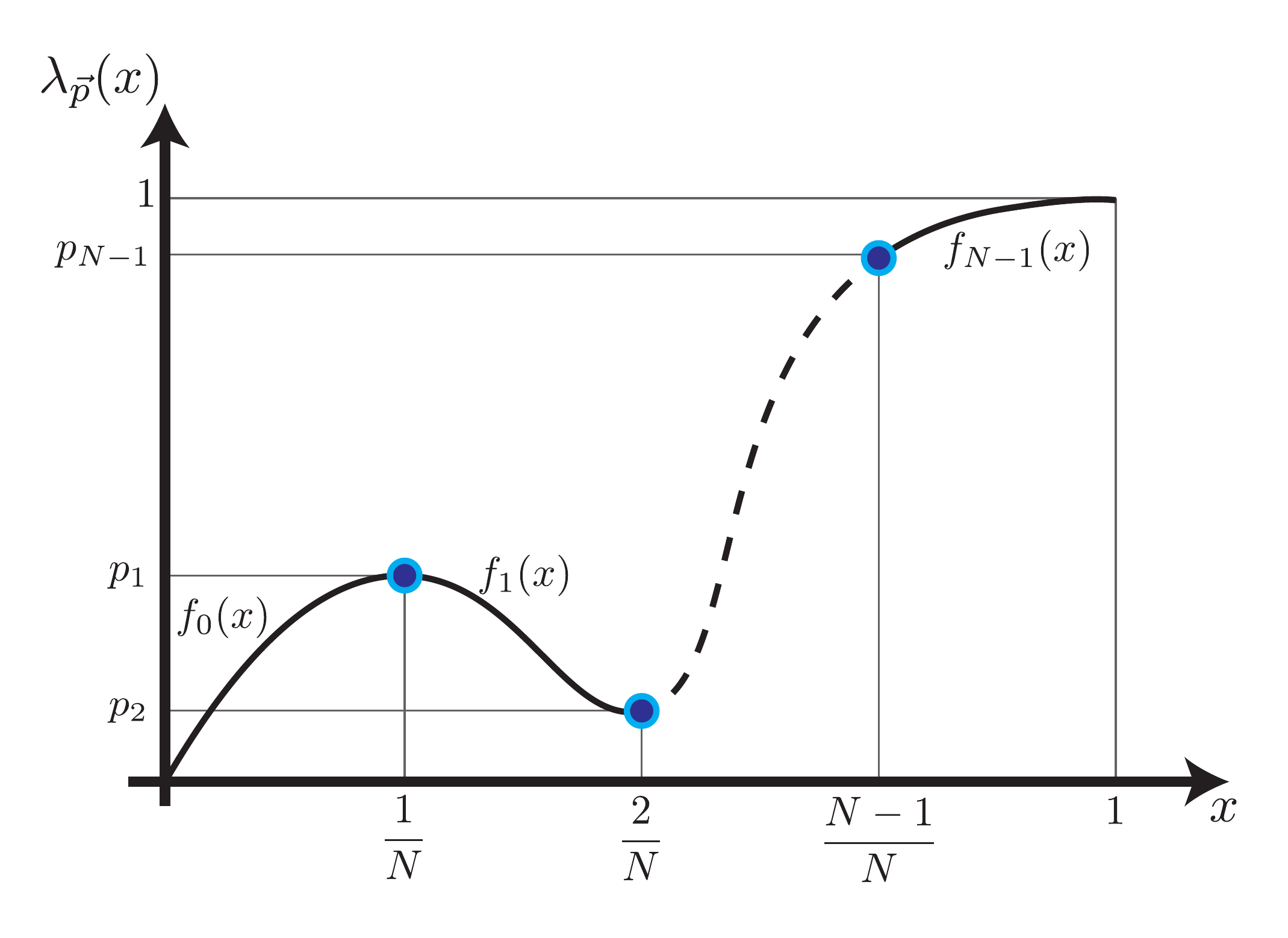}
\caption{Parametrized schedule function $\lambda_{\vec{p}}(x)$. The parameters $p_j$ define the points $(j/N, p_j)$, which are interpolated using a piecewise monotonic cubic interpolation.}
\label{Fig01}
\end{figure}

In a similar form, we define the analog QAOA (AQAOA) as the optimization of the continuous-time evolution given by Eq.~(\ref{Eq02}). To do this, we propose a general parametrization of the schedule function $\lambda (t/T)$ using fixed points whose coordinates are variational parameters to be optimized, and we use a piece-wise cubic interpolator through the points to draw the schedule function (see figure \ref{Fig01}). Using such parametrization in principle, we can cover all the possible schedule functions by adding more parameters (points). To obtain enough conditions for the piece-wise cubic interpolator, we consider continuity in the schedule function and its first derivate, as well as monotonicity, to ensure that a bound in the parameters also means a bound in the function.

These three conditions define a cubic Hermite spline. As this interpolation ensures the continuity of the function and its first derivative, the resulting function can be implemented in an analog way, and any experimental restriction that limits the possible values of the schedule function can be treated as bounds in the parameters. We can note that other interpolation fails in these essential conditions. For example, a high-degree polynomial can introduce unnecessary oscillations, a piece-wise linear interpolator yields functions with discontinuous first derivative, which is harder to implement, and a piece-wise quadratic interpolator cannot ensure monotonicity between two adjacent points, which could return values for the schedule function that go beyond the experimentally implementable range.

The core of our AQAOA algorithm is the construction of the parametrized schedule function of the time-dependent Hamiltonian in Eq.~(\ref{Eq01}). As mentioned above, it is defined as a piece-wise function as follows

\begin{eqnarray}
\lambda_{\vec{p}}(x)=f_k(x) ; \quad \frac{k}{N}\le x \le \frac{k+1}{N} \, ;
\end{eqnarray}
\begin{eqnarray}
f_k(x)=a_kx^3+b_kx^2+c_kx+d_k ; \quad && f_k(k/N)=p_k,\nonumber\\
&&f_k(k+1/N)=p_{k+1} \, ;
\end{eqnarray}
and ${\vec{p}}=\{p_0,p_1,\dots,p_{N-1},p_N\}$. It means that $\lambda_{\vec{p}}(x)$ is given by a cubic interpolation function of the points ($k/N,p_k$), with $k=\{0,\dots,N\}$ and $p_0=0$ and $p_N=1$. Now we have the following unitary evolution
\begin{equation}
U_{\vec{p}}(t)=\mathcal{T}e^{-i\int_0^t\{[1-\lambda_{\vec{p}}(\tau/T)]H_{i}+\lambda_{\vec{p}}(\tau/T)H_{f}\}d\tau},
\end{equation}
and the cost function
\begin{equation}
E(\vec{p})=\bra{\phi}U_{\vec{p}}^{\dagger}H_fU_{\vec{p}}\ket{\phi} \, .
\label{Eq10}
\end{equation}
Here, $U_{\vec{p}}\equiv U_{\vec{p}}(t)$, and $\ket{\phi}$ is the ground state of the initial Hamiltonian $H_i$. We need to mention that $H_i$ and $H_f$ can be the Hamiltonians that quantum annealers can implement currently, where our requirement is only the manipulation of the schedule function in the form of a cubic interpolation function. We note that the total evolution time $T$ plays the role of the circuit depth. In this case, it is independent of the number of parameters to optimize, which is a fundamental difference from the digital QAOA. We also require this evolution to be coherent, such that $T$ needs to be smaller than the coherence time of the quantum annealer. 

There is an important distinction between our proposed AQAOA and the digital QAOA. Both approaches are heuristic algorithms, but the former is suitable to be implemented in an analog quantum computer such as a quantum annealer, while the latter is a paradigm for gate-base quantum computers. Therefore, the sources of error also differ in these two approaches. For digital QAOA, all the errors due to the digitalization of the evolution are related to the maximal circuit size allowed by the quantum computer. Whereas the error sources in AQAOA are related to the total evolution time $T$, which is limited by the coherence time of the quantum device and to the number of parameters used to build and optimize the schedule function. Since the schedule function is defined piece-wise, the number of piece-wise functions involved in the optimization is proportional to the number of parameters considered.

Finally, we need to mention that our AQAOA algorithm is not a shortcut to adiabaticity (STA) technique~\cite{GueryOdelin2019RevModPhys}, even if, in the last year, STA techniques have been used in gate-based quantum computers to minimize the energy of a problem Hamiltonian surpassing previous paradigms~\cite{Hegade2021PhysRevA, Hegade2021PhysRevAppl}. In STA protocols, the evolution mimics an adiabatic path, which means that STA protocols follow the instantaneous eigenstate of a given Hamiltonian. In our proposal, this is not the case. We propose a heuristic algorithm that can be defined as a variational quantum algorithm suitable for quantum annealers. Our work is focused on obtaining a final state that minimizes the energy in the final Hamiltonian and not on following a given evolution.

In the next section, we will test our AQAOA numerically for a cubic interpolation for different problem Hamiltonians~$H_f$. We need to mention that we do not consider $0<p_j<1$. As in the digital version of QAOA, the parameters are completely free and only bound by the experimental setup limitations. Then, we allow that $\lambda_{\vec{p}}(x)>1$ or $\lambda_{\vec{p}}(x)<0$, where any bound in the parameters depends on the experimental capabilities, and it is not a theoretical limitation. The unbound schedule function is not a problem because our method is inspired by the adiabatic evolution but does not follow, in general, an adiabatic path. We highlight that our AQAOA only focuses on minimizing the cost function and does not care about the distance to the adiabatic evolution path. Finally, we request that the interpolation be soft, which means continuity in the first derivative, and monotonic to control the maximal and minimal values of the function; that is, the second derivative will not change of sign between two adjacent points.

\section{Numerical results}
The performance of our algorithm is calculated using the relative error of the energy obtained by our AQAOA, which means
\begin{equation}
\epsilon_R=\left|\frac{\bra{\phi}U_{\vec{p}}^{\dagger}H_fU_{\vec{p}}\ket{\phi}-\bra{\psi}H_f\ket{\psi}}{\bra{\psi}H_f\ket{\psi}}\right|,
\end{equation}
where $\ket{\psi}$ is the ground state of the final Hamiltonian $H_f$. Also, we can consider the fidelity
\begin{equation}
 F=\abs{\bra{\psi}U_{\vec{p}}\ket{\phi}}^2.
 \end{equation}
Nevertheless, the last could fail as a good performance measure for degenerate ground states because our algorithm only focuses on minimizing the energy and does not care about a specific ground state. 

For all our examples, we will use the following initial Hamiltonian
\begin{equation}
H_i=\frac{\omega_i}{2}\sum_{j=1}^n\sigma_x^{(j)} \, ,
\end{equation}
where $\sigma_x^{(j)}$, is the $x$-Pauli matrix of the $j$th qubit, $\omega_i$ is the initial frequency gap for all the qubits, and $n$ is the total number of qubits.
 \begin{figure}[t]
\centering
\includegraphics[width=0.8\linewidth]{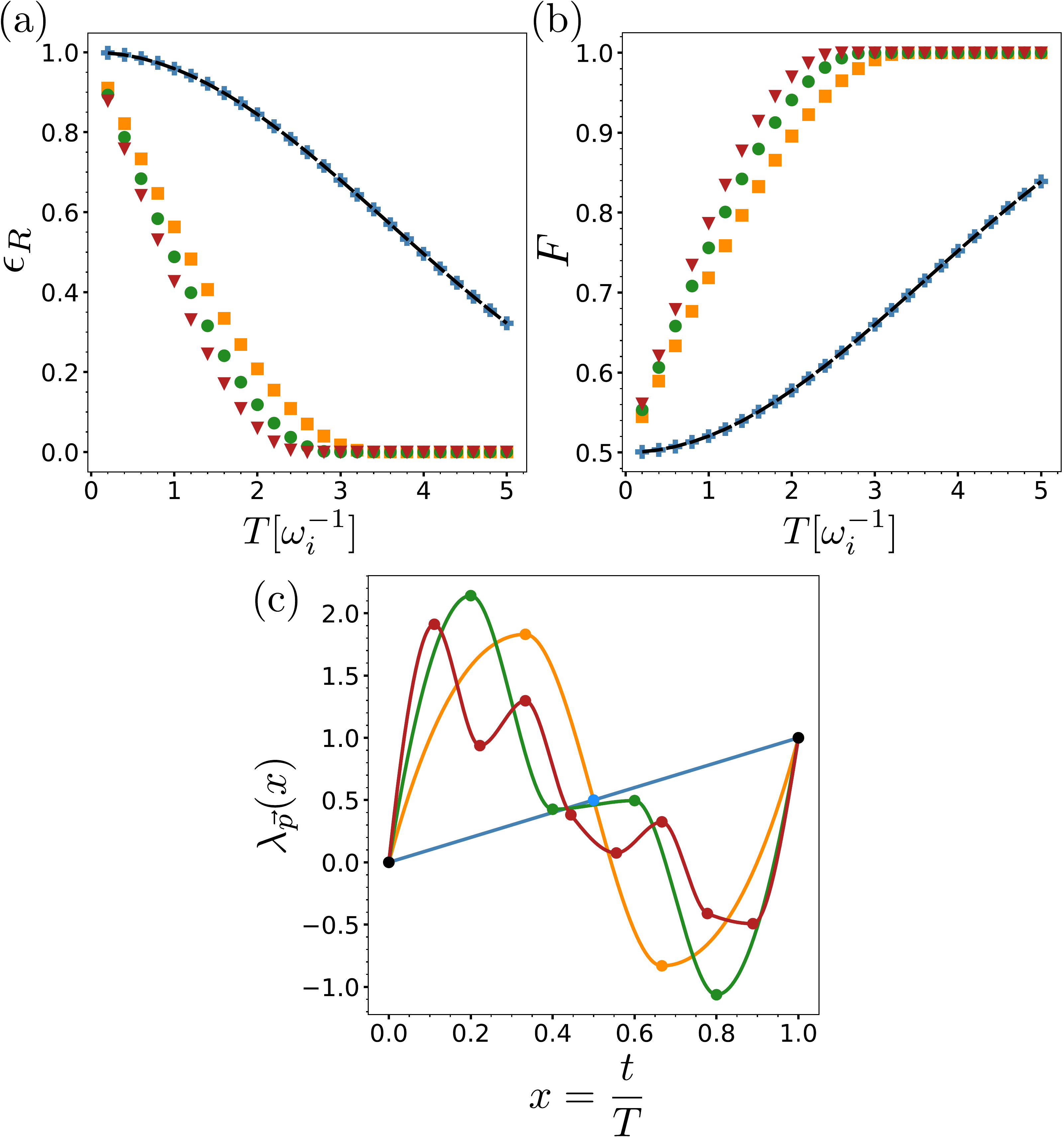}
\caption{Performance of the AQAOA for the one-qubit case given by Hamiltonian of Eq.~(\ref{Eq14}). The horizontal axis is the total time $T$ considered for the algorithm in units of $\omega^{-1}$. We calculate the relative error (a) and the fidelity (b), considering one (blue crosses), two (orange squares), four (green dots), and eight (red triangles) parameters for the algorithm. The black dashed line shows the performance using a linear schedule function. Figure (c) shows the schedule function for one (blue line), two (orange line), four (green line) and eight (red line) free parameters (dots) for $T=3$.}
\label{Fig02}
\end{figure}

\subsection{One-qubit case}
For our first example, we consider the trivial case of a single qubit final Hamiltonian
\begin{equation}
H_f=\frac{\omega_f}{2}\sigma_z^{(1)}.
\label{Eq14}
\end{equation}
For simplicity, we consider the same frequency for the initial and final Hamiltonian ($\omega_i=\omega_f$). The performance of our algorithm is collected in Fig.~\ref{Fig02}a for the relative error $\epsilon_R$, and Fig.~\ref{Fig02}b for the fidelity. Each point (marker) represents the performance of the algorithm for different total algorithmic time $T$ and a different number of parameters in the minimization process, namely, one parameter (blue crosses), two parameters (orange squares), four parameters (green dots), and eight parameters (red triangles). Moreover, we compare the performance of our algorithm with adiabatic quantum computing using the most common schedule function, i.e., the linear schedule function $\lambda(t)=t/T$ (black dashed line). From these two figures, we can see that a time of $T=4~[\omega_i^{-1}]$ is enough for our algorithm to find the solution to the problem with only two parameters, while the linear function needs approximately double time. On the other hand, the algorithm performance using one parameter for the optimization process follows the same performance as the linear schedule function. If we consider a superconducting flux qubit $\omega_i\sim2$~[GHz], then for a time $T\sim2$~[ns], our algorithm produces the correct solution using two parameters for a single qubit Hamiltonian. Finally, Fig.~\ref{Fig02}c shows optimal schedule functions for one (blue), two (orange), four (green), and eight (red) free parameters. We point out that the endpoints (0.0) and (1,1) are fixed. The oscillations in the schedule functions mean that we are not following an adiabatic evolution, which is also observed by the fact that we have zones where $\lambda_{\vec{p}}(x)>1$ and $\lambda_{\vec{p}}(x)<0$. It implies that the adiabatic theorem does not constrain the time for our final result.
 
\subsection{Hydrogen Molecule}
\begin{figure}[t]
\centering
\includegraphics[width=0.8\linewidth]{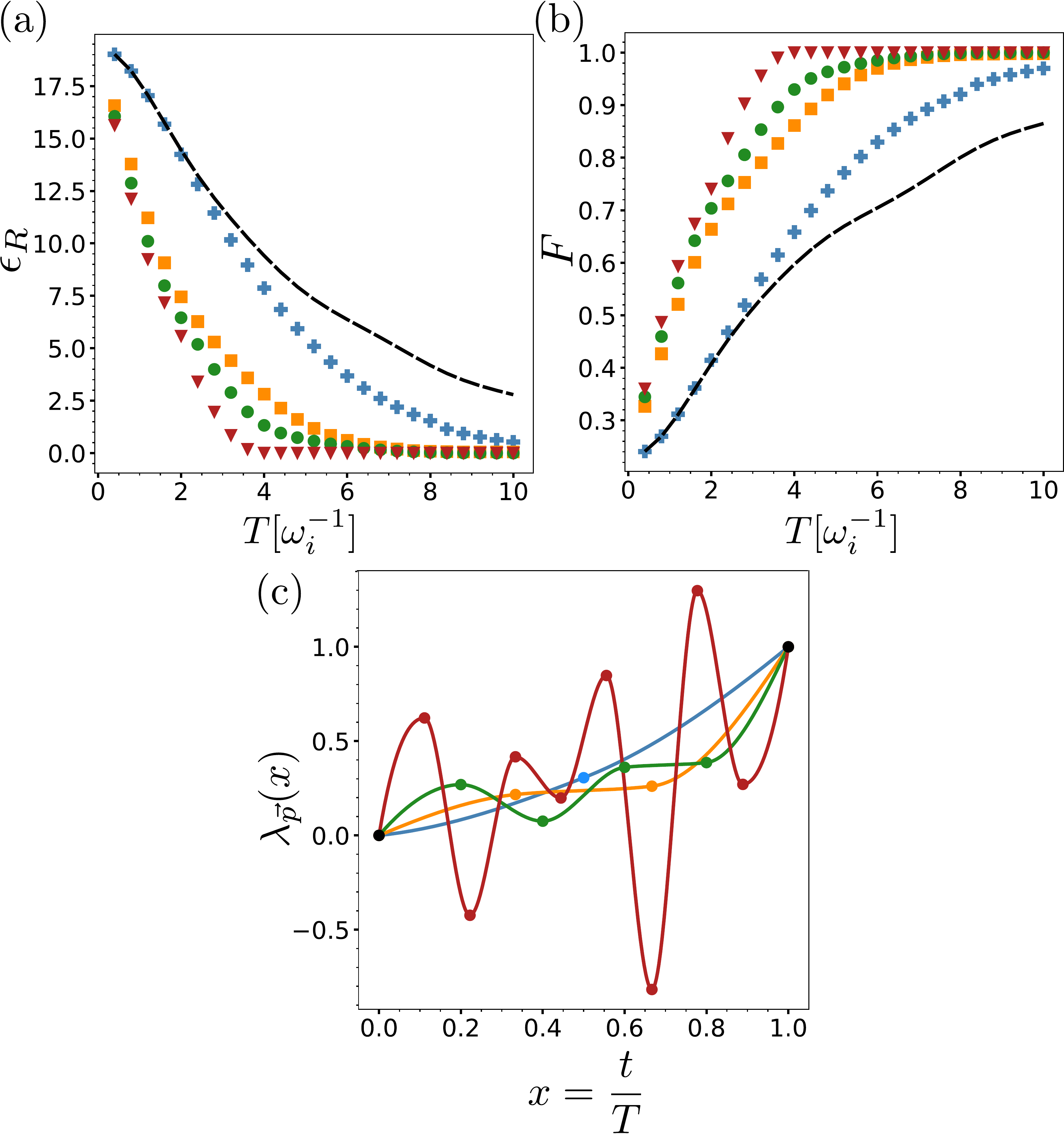}
\caption{Performance of the AQAOA for the Hydrogen-molecule case given by the Hamiltonian of Eq.~(\ref{Eq15}). The horizontal axis is the total time $T$ considered for the algorithm in units of $\omega^{-1}$. We calculate the relative error (a) and the fidelity (b), considering one (blue crosses), two (orange squares), four (green dots), and eight (red triangles) parameters for the algorithm. The black dashed line shows the performance using a linear schedule function. Figure (c) shows the schedule function for one (blue line), two (orange line), four (green line) and eight (red line) free parameters (dots) for $T=5$.}
\label{Fig03}
\end{figure}

Our next example is a non-stoquastic Hamiltonian describing a hydrogen molecule~\cite{OMalley2016PRX} with a bond length of $0.2\,[\AA]$
\begin{equation}
H_f=g_0\mathbb{I}+g_1\sigma_z^{(1)}+g_2\sigma_z^{(2)}+g_3\sigma_z^{(1)}\sigma_z^{(2)}+g_4\sigma_y^{(1)}\sigma_y^{(2)}+g_5\sigma_x^{(1)}\sigma_x^{(2)},
\label{Eq15}
\end{equation}
with $g_0=2.8489$, $g_1=0.5678$, $g_2=-1.4508$, $g_3=0.6799$, and $g_4=g_5=0.0791$. The performance of our algorithm is shown in Fig.~\ref{Fig03}a for the relative error $\epsilon_R$ and Fig.~\ref{Fig03}b for the fidelity. In this case, we consider the same number of parameters as in the previous case, and again we compare the performance of the algorithm with the performance using the linear schedule function, black dashed line in the figure. From these two figures, we can see that for a time of $T=8~[\omega_i^{-1}]$; the AQAOA algorithm can find the solution to the problem with only two parameters with a fidelity larger than $0.99$. In this case, the performance of the algorithm using only one parameter for the optimization process improves the performance of the linear schedule function, obtaining fidelities over $0.95$ for a time $T=10~[\omega_i^{-1}]$. Considering a quantum annealer based on flux qubits, our algorithm gets the correct solution using two parameters for $T\sim4$~[ns]. Finally, Fig.~\ref{Fig03}c shows optimal schedule functions for a different number of parameters.

\subsection{Ising and Heisenberg Hamiltonians}
\begin{figure}[t]
\centering
\includegraphics[width=0.8\linewidth]{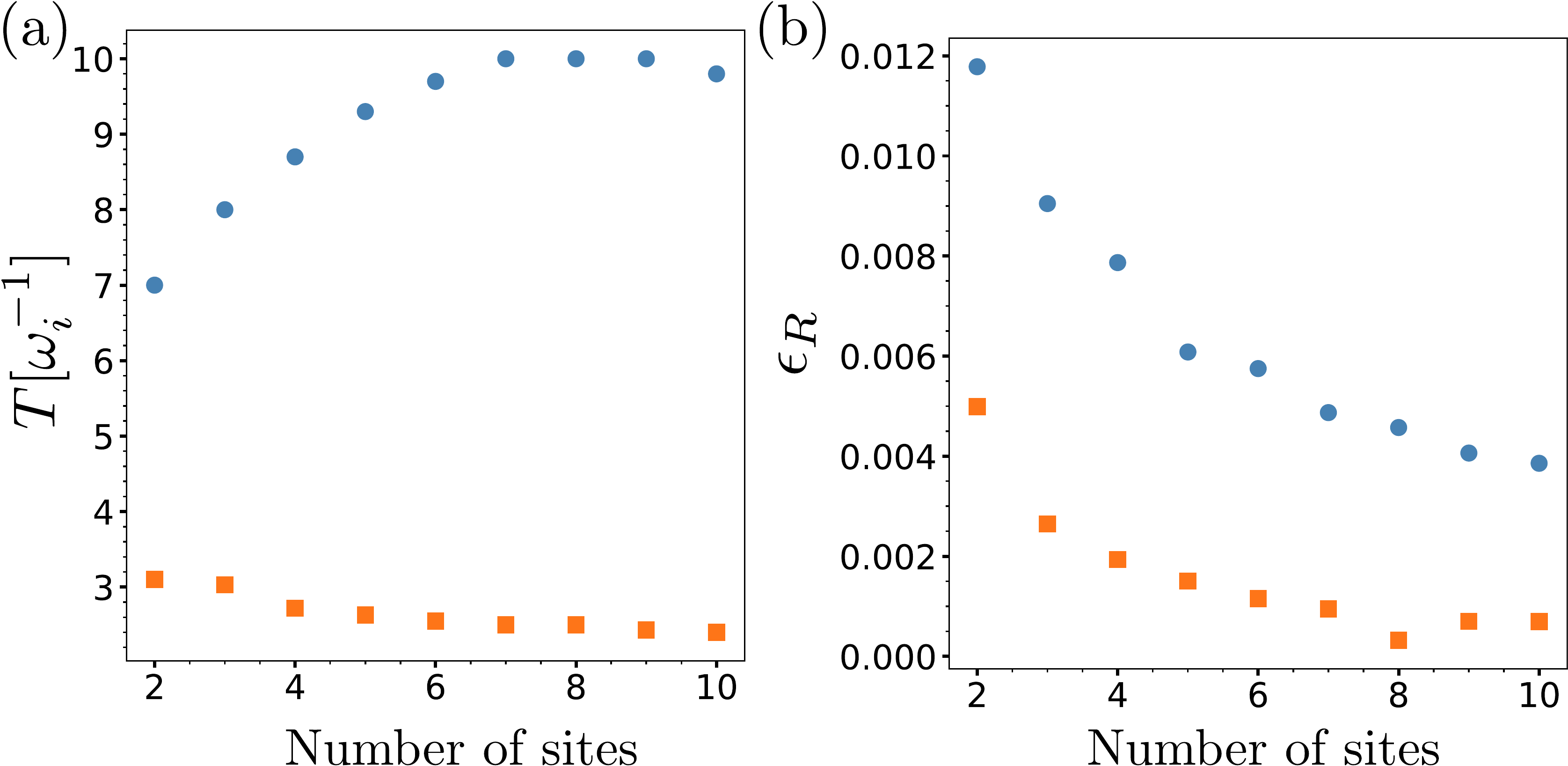}
\caption{Performance of the AQAOA for an Ising chain (blue dots) and homogeneous Heisenberg chain (orange squares) with a different number of sites. (a) Total running time $T$ necessary for fidelity over $0.99$ using the same number of parameters as sites in the chain. (b) The relative error $\epsilon_R$ for the same cases than in (a).}
\label{Fig04}
\end{figure}
Finally, we consider the scaling in the total algorithmic time $T$ for two kinds of nearest-neighbor Hamiltonians. First, we consider the stoquastic Hamiltonian of an Ising chain given by
\begin{equation}
H_f=\frac{1}{2}\omega_f\sum_{k=1}^{N}\sigma_z^{(k)}- J\sum_{k=1}^{N-1}\sigma_z^{(k)}\sigma_z^{(k-1)},
\end{equation}
and second, the non-stoquastic Hamiltonian of a homogeneous Heisenberg chain which reads
\begin{eqnarray}
H_f=&&\frac{1}{2}\omega_f\sum_{k=1}^{N}\sigma_z^{(k)}\nonumber\\
&&- J\sum_{k=1}^{N-1}\left(\sigma_x^{(k)}\sigma_x^{(k-1)}+\sigma_y^{(k)}\sigma_y^{(k-1)}+\sigma_z^{(k)}\sigma_z^{(k-1)}\right).
\end{eqnarray}

For simplicity we consider in both cases $\omega_i=\omega_f=2J$. We use $\mathcal{N}$ parameters for the optimization process for the chain of $\mathcal{N}$ sites. This means that if we consider a chain of $\mathcal{N}=5$ sites for the Ising or Heisenberg Hamiltonians, we use $5$ parameters in the parametrization of the schedule function. Furthermore, we are interested in the time $T$ needed to get a solution with fidelity over $0.99$. These results are collected in Fig.~\ref{Fig04}a, which shows $T$ for an Ising chain (blue dots) and homogeneous Heisenberg chain (orange dots) as a function of the number of sites $\mathcal{N}$. Moreover, Fig.~\ref{Fig04}b shows the corresponding relative error $\epsilon_R$ for the same cases as in Fig.~\ref{Fig04}a. We note that for the homogeneous Heisenberg chain, the time $T$ decreases with the number of qubits in the system. This suggests that with a time $T=3~[\omega_i^{-1}]$, the algorithm can obtain results with fidelities larger than $0.99$ with $N$ parameters for a chain of $N$ sites. On the other hand, the time $T$ for the Ising chain increases with the number of sites and reaches a maximum value in $N=8$ sites with a time $T\sim10~[\omega_i^{-1}]$. Finally, in a superconducting circuit platform, an advanced architecture for quantum computing, the frequency $\omega_i$ is in the order of a few GHz. It implies that for most superconducting circuit setups $\omega_i>1$~[GHz]$~\Rightarrow \omega_i^{-1}<1$~[ns] while considering typical coherent times in the scale of microseconds, it means larger than $10^3\omega_i^{-1}$. Therefore, our AQAOA algorithm gives an interesting approach to achieving fast and high-fidelity approximations for optimization problems suitable for quantum annealers in a coherent manner.

Finally, we remark that for AQAOA, as in any variational algorithm, the classical optimization process can be challenging as the size of the parameter space increases, where the optimization landscape is more likely to be afflicted by barren plateaus~\cite{Uvarov2021JPhysA} and many local minima. This is a well-known problem of black-box optimization algorithms used in variational approaches, that is, an optimization process without previous knowledge about the function to be optimized~\cite{Abell2013Book,Kvasov2015AdvEngSoft,Parnianifard2018DecSciLett,Munoz2015InfSci}. This suggests that an algorithmic study of the different optimization strategies (classical algorithms) for different Hamiltonians and structures is necessary to enhance these heuristic algorithms, that is, to understand which is the more convenient optimizer based on the Hamiltonian structure, and developing an appropriate research field to merge computer science and quantum computing.

\section{Conclusions}
This work introduced an analog version of the quantum approximate optimization algorithm suitable for current quantum annealers. Our algorithm is based on a general parametrization of the schedule function, which produces any function if we consider enough parameters. We can coherently produce fast and high-fidelity solutions by optimizing this parametrization in the same way as in the standard QAOA algorithm. We test our algorithm numerically for different cases. First, for a single qubit Hamiltonian; second, for the ground state energy of the Hydrogen molecule; and third, for an Ising and homogeneous Heisenberg chain with a different number of sites. The latter ranges from $2$ to $10$, obtaining high fidelities with a relatively low number of parameters in all studied cases, amounting to the same number of parameters as the qubits in the Hamiltonian.

Furthermore, a possible experimental implementation of our AQAOA depends on two features: first, a quantum annealer capable of producing the desired final or problem Hamiltonian $H_f$, and second, a quantum annealer with a schedule function that can be manipulated. The experimental limitations on the final Hamiltonian will determine the classes of problems we can solve. Also, depending on the experimental manipulability of the schedule function, our optimization process will require more or fewer constraints.

Finally, this work paves the way for efficient implementations of optimization algorithms in analog devices such as current quantum annealers, exploiting the inherent quantum nature of the device.

\section{Acknowledgments}

The authors acknowledge support from projects STCSM (2019SHZDZX01-ZX04 and 20DZ2290900), Junta de Andaluc\'ia (P20-00617), ANID Subvenci\'on a la Instalaci\'on en la Academia SA77210018, and ANID Proyecto Basal AFB 180001.\\

\end{document}